\documentclass[final]{ws-procs9x6}
\usepackage{upgreek}

\begin{document}

\title{Testing Thermo-acoustic Sound Generation in Water with Proton and Laser
  Beams\footnote{\uppercase{T}his work is supported by the \uppercase{G}erman \uppercase{BMBF}
    \uppercase{G}rant \uppercase{N}o. 05~\uppercase{CN}2\uppercase{WE}1/2.}}
\author{K.~Graf, G.~Anton, J.~H{\"o}ssl, A.~Kappes, T.~Karg, U.~Katz, R.~Lahmann, C.~Naumann,
  K.~Salomon and C.~Stegmann}
\address{Physikalisches Institut,\\ Friedrich-Alexander-Universit{\"a}t Erlangen-N{\"u}rnberg,\\
  Erwin-Rommel-Stra{\ss}e 1,\\ 91058 Erlangen, Germany\\ E-mail: kay.graf@physik.uni-erlangen.de}
\maketitle 

\abstracts {Experiments were performed at a proton accelerator and an infrared laser facility to
  investigate the sound generation caused by the energy deposition of pulsed particle and laser
  beams in water. The beams with an energy range of $1\,\mathrm{PeV}$ to $400\,\mathrm{PeV}$ per
  proton beam spill and up to $10\,\mathrm{EeV}$ for the laser pulse were dumped into a water volume
  and the resulting acoustic signals were recorded with pressure sensitive sensors. Measurements
  were performed at varying pulse energies, sensor positions, beam diameters and temperatures. The
  data is well described by simulations based on the thermo-acoustic model. This implies that the
  primary mechanism for sound generation by the energy deposition of particles propagating in water
  is the local heating of the media giving rise to an expansion or contraction of the medium
  resulting in a pressure pulse with bipolar shape.  A possible application of this effect would be
  the acoustical detection of neutrinos with energies greater than 1 EeV.}

\section{Introduction}

The production of hydrodynamic radiation (ultrasonic pressure waves) by fast particles passing
through liquids was first predicted already in 1957 leading to the development of the so-called
thermo-acoustic model in 1979\cite{Askariyan1,Askariyan2}. The model allowed to describe the primary
production mechanism of the bipolar shaped acoustic signals measured in an experiment with proton
pulses in fluid media\cite{Sulak}. According to the model, the energy deposition of particles
traversing liquids leads to a local heating of the medium which can be regarded as instantaneous
with respect to the hydrodynamic time scale. Due to the temperature change the medium expands or
contracts according to its volume expansion coefficient $\alpha$. The accelerated motion of the
heated medium forms an ultrasonic pulse which propagates in the volume.  The wave equation
describing the pulse is\cite{Askariyan2}
\begin{equation}
  \Delta p(\vec{r},t) - \frac{1}{c_s^2} \cdot \frac{\partial^2 p(\vec{r},t)}{\partial t^2} =-
  \frac{\alpha}{C_p} \cdot \frac{\partial^2 \epsilon(\vec{r},t)}{\partial t^2}
\label{waveequation}
\end{equation}
and can be solved using the Kirchhoff integral:
\begin{equation}
 p(\vec{r},t) = \frac{1}{4\pi}\frac{\alpha}{C_p}\int_V \frac{\mathrm{d}V'}{|\vec{r} -
 \vec{r}\,'|}\;\frac{\partial^2}{\partial t^2}\,\,\epsilon\left(\vec{r}\,',t -\frac{|\vec{r} - \vec{r}\,'|}{c_s}\right)\textrm{.}
\label{pressureequation}
\end{equation}
Here $p(\vec{r},t)$ denotes the hydrodynamic pressure at a given place and time, $c_s$ the speed of
sound in the medium, $C_p$ its specific heat capacity and $\epsilon(\vec{r},t)$ the energy
deposition density of the particles. The resulting pressure field is determined by the spatial and
temporal distribution of $\epsilon$ and by $c_s$, $C_p$ and $\alpha$, the latter three depending on
the temperature. A controlled variation of these parameters in laboratory experiments and a study of
the resulting pressure signals allows therefore a precise test of the thermo-acoustic model. One
decisive test is the disappearance of the signal at $4^{\circ}\mathrm{C}$ in water, the medium
considered in the following, due to the vanishing $\alpha$ at this temperature.

However, in previously conducted experiments investigating this effect in different liquids, the
observed pulses could not be unambiguously verified as
thermo-acoustical\cite{Sulak,Hunter1,Hunter2,Albul}. There the variation of the pulse amplitude with
the temperature in water showed not the predicted dependency and particularly not the
disappearance of the signal at 4$^{\circ}$C.

\section{Conducted Experiments}

The experiments presented in this paper were performed with a pulsed $1064 \, \mathrm{nm}$ Nd:YAG
laser facility located at our institute, and the $177 \, \mathrm{MeV}$ proton beam of the ``Gustaf
Werner Cyclotron'' at the ``Theodor Svedberg Laboratory'' in Uppsala, Sweden.  The beams were dumped
into a $150 \times 60 \times 60 \, \mathrm{cm}^3$ water tank, where the acoustic field was measured
with several position-adjustable hydrophones (pressure sensitive sensors based on the piezo-electric
effect). The temperature of the water was varied between $1^\circ \mathrm{C}$ and $20^\circ
\mathrm{C}$ with a precision of $0.1^\circ \mathrm{C}$ by cooling and gradual controlled homogeneous
reheating of the whole water volume.

The spill energy of the proton beam was varied from $10 \, \mathrm{PeV}$ to $400 \, \mathrm{PeV}$,
the beam diameter was approx.~$1 \, \mathrm{cm}$ and the spill time $30 \, \mathrm{\upmu s}$. For
$177\,\mathrm{MeV}$ protons, the energy deposition in the water along the beam axis ($z$-axis, beam
entry into the water at $z=0\,\mathrm{cm}$) is relatively uniform up to $z=20\,\mathrm{cm}$ ending
in the prominent Bragg-peak at $z \approx22\,\mathrm{cm}$. The laser pulse energy was adjusted
between $0.1 \, \mathrm{EeV}$ and $10 \, \mathrm{EeV}$ at a beam diameter of a few mm, the pulse
length was fixed at $9 \, \mathrm{ns}$. The laser energy density deposited along the beam axis had
an exponential decrease with a absorption length of $(6.0\pm0.1) \, \mathrm{cm}$.  The two
experiments enabled us to use different spatial and temporal distributions of the energy deposition
as well as two different kinds of energy transfer into the medium, i.e. by excitation by both beams
and additionally by ionisation by the proton beam.

The sensors used were characterised and found to be linear in amplitude response, the frequency
response was flat up to the main resonance at $50\,\mathrm{kHz}$ with a sensitivity of
approx.~$-150\,\mathrm{dB\,re\,}1\mathrm{V}/\upmu\mathrm{Pa}$. The sensitivity dependence on
temperature was measured and the relative decrease was found to be less than $1.5\%$ per
$1^{\circ}\mathrm{C}$.

For every set of experimental parameters the signals of 1000 beam pulses were recorded, with a
sampling rate of $10\,\mathrm{MHz}$, sufficient for the typical frequency range of the signals of
$5\,\mathrm{kHz}$ to $100\,\mathrm{kHz}$.
\section{Results}

The measured bipolar signals were found to be in good agreement with simulations based on the
thermo-acoustic signal generation mechanism. The hydrodynamic nature of the signal was proven by
determining the propagation times of the signals at different positions in the $x$-direction
perpendicular to the beam axis. They were consistent with the expected propagation times for sound
in water.  Also the investigated signal dependencies on beam energy, beam width and sensor distance
from the beam show very good agreement with the simulation based on the thermo-acoustic model.

\begin{figure}[ht]
  \centerline{\epsfxsize=4.4in\epsfbox{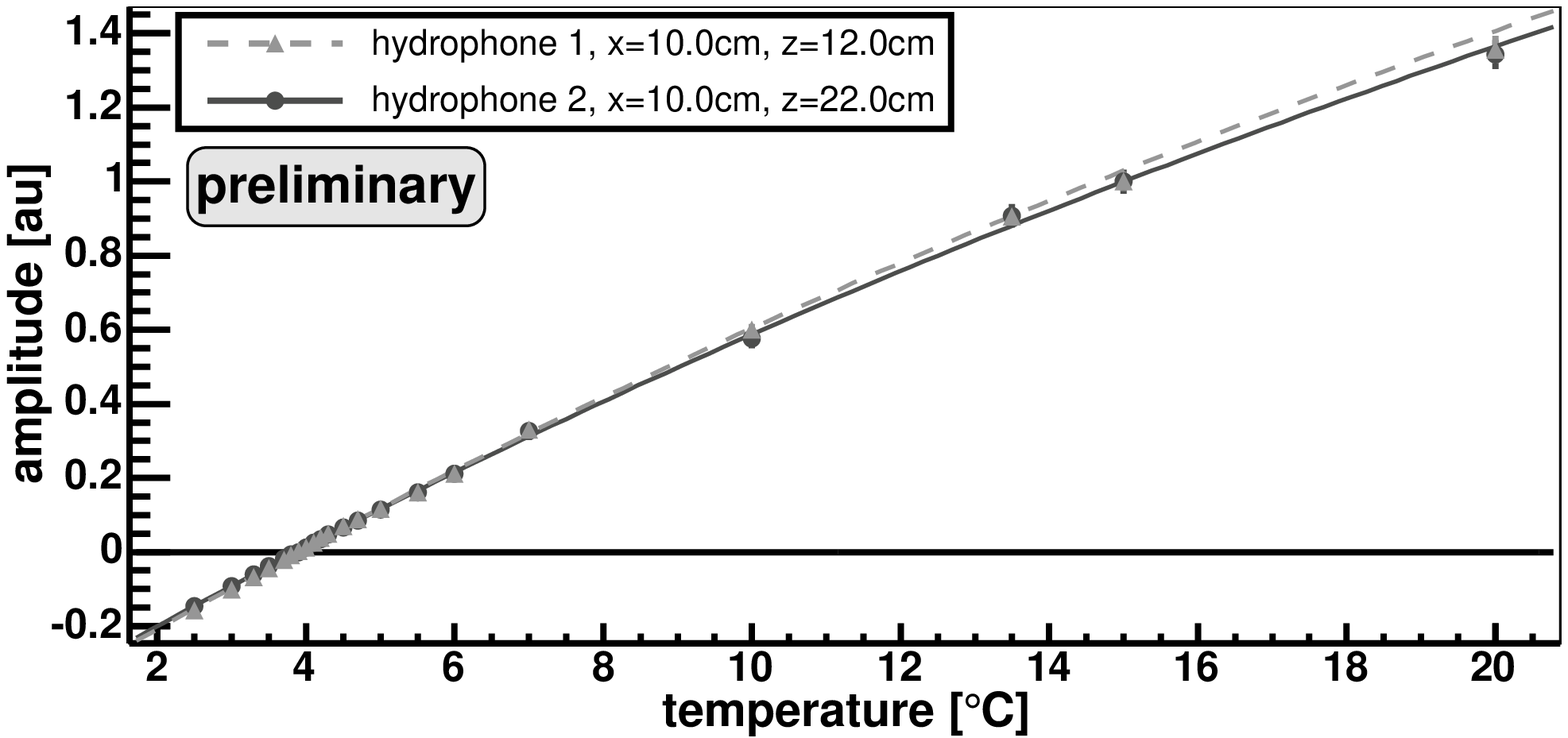}}
\caption{Measured signal amplitude of the bipolar acoustic signal produced by laser pulses at different
  temperatures fitted with the model expectation as described in the text. All amplitudes were
  normalised to $1$ at $15.0^{\circ}\mathrm{C}$.\label{fig1a}}
\vspace{5pt}
\centerline{\epsfxsize=4.4in\epsfbox{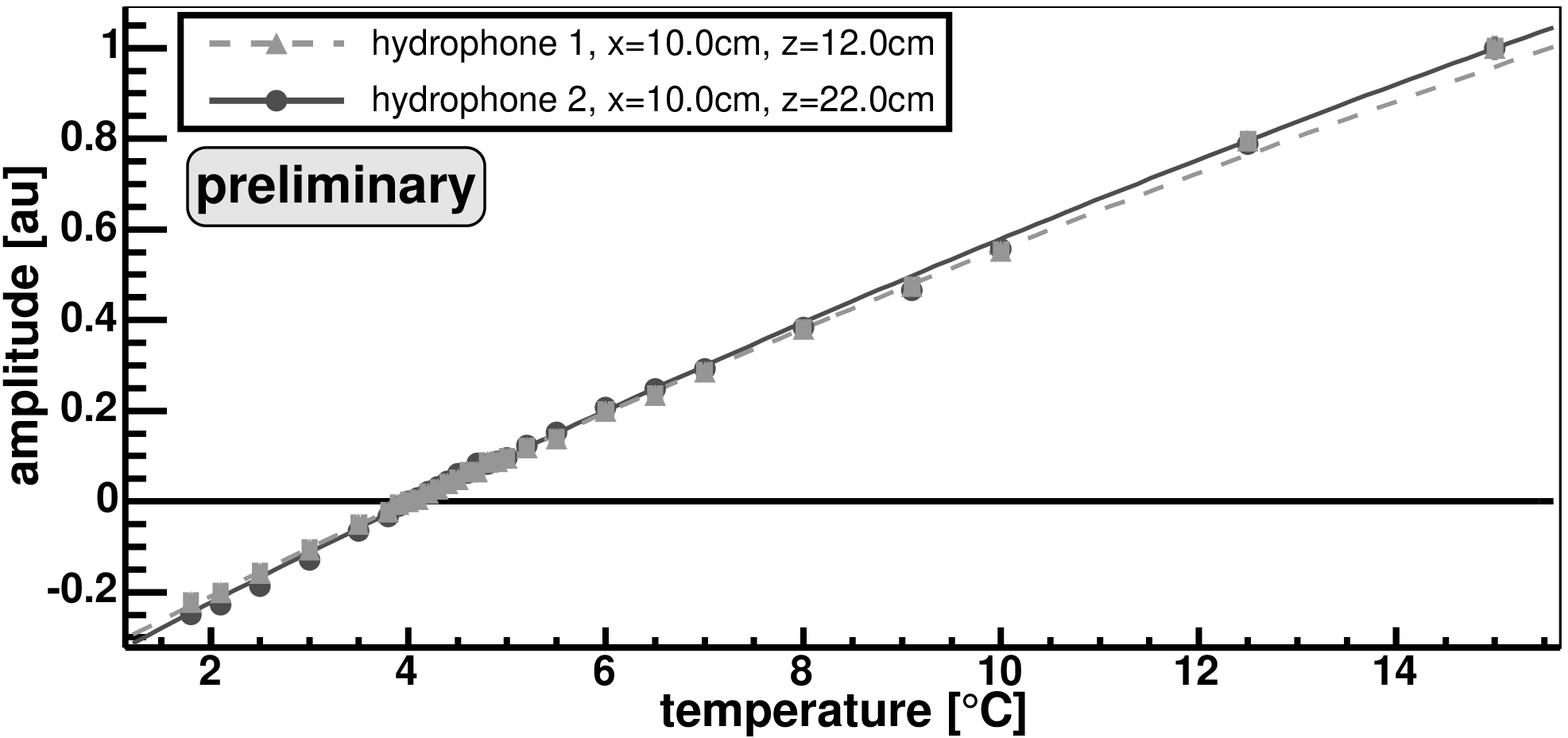}}
\caption{Measured signal amplitude of the bipolar acoustic signal produced by proton pulses at different
  temperatures fitted with the model expectation as described in the text. The non-thermo-acoustic
  signal at 4.0$^{\circ}$C was subtracted at every temperature. The amplitudes were afterwards
  normalised to $1$ at $15.0^{\circ}\mathrm{C}$.\label{fig1b}}
\end{figure}

Figures~\ref{fig1a} and \ref{fig1b} show the temperature dependence of the peak-to-peak amplitude of
the bipolar signals for the two experiments, where a positive (negative) sign denotes a leading
positive (negative) peak of the signal. The two data sets shown in each figure were recorded by two
sensors positioned at $x=10\,\mathrm{cm}$ perpendicular to the beam axis and at $z=12\,\mathrm{cm}$
and $z=22\,\mathrm{cm}$ along the beam axis, respectively. In the case of the proton beam setup the
hydrophone positions correspond roughly to the $z$-position of the Bragg-peak and a $z$-position
half way between the Bragg-peak and the beam entry into the water, respectively. For comparability,
the same positions and the same sensors were chosen for the laser experiment.
\newpage 

The laser beam signal shown in Fig.~\ref{fig1a} changes its polarity around $4 ^\circ \mathrm{C}$,
as expected from the thermo-acoustic model. The model expectation for the signal amplitude, which is
proportional to $\alpha/C_p$ and vanishes at $4^\circ\mathrm{C}$, is fitted to the experimental
data. In the fit an overall scaling factor and a constant temperature shift were left free as fit
parameters.  The fit yielded a zero-crossing of the amplitude at $(3.9 \pm 0.1)^\circ \mathrm{C}$,
where the error is dominated by the systematic uncertainty in the temperature setting. The
zero-crossing is in good agreement with the expectation of $4.0 ^\circ \mathrm{C}$.  Analysing the
proton data in the same way yielded a shape slightly deviating from the model expectation, and a
zero-crossing significantly different from $4.0^\circ \mathrm{C}$ at $(4.5 \pm 0.1)^\circ \mathrm{C}$.
In view of the results from the laser beam measurements, we subtracted the residual signal at
$4.0^\circ \mathrm{C}$, which has an amplitude of approx.~$5\%$ of the $15.0^{\circ}\mathrm{C}$
signal, from all signals, assuming a non-temperature dependent effect on top of the thermo-acoustic
signal.  The resulting amplitudes shown in Fig.~\ref{fig1b} are well described by the model
prediction.

The source of the small non-thermo-acoustic signal which was only seen in the proton experiment
could not be unambiguously verified with these experiments. The obvious difference to the laser
experiment are the charges involved both from the protons themselves and the ionisation of the
water.  For clarification further experiments are needed either with ionising neutral particles
(e.g.~synchrotron radiation) or with charged particles (e.g.~protons, $\upalpha$-particles) with
more sensors positioned around the Bragg-peak. With such experiments it might be possible to
distinguish between the effect of ionisation in the water and of net charge introduced by charged
particles.

\section{Conclusions}
We have demonstrated that the sound generation mechanism of intense pulsed beams is well described
by the thermo-acoustic model. In almost all aspects investigated, the signal properties are
consistent with the model. Relying on the model allows to calculate the characteristics of sound
pulses generated in the interaction of high energy particles in water with the input of the energy
deposition of the resulting cascade. A possible application of this technique would be the
detection of neutrinos with energies $\gtrsim 1\,\mathrm{EeV}$\cite{Karg}.


\begin{thebibliography}{0}
  
\bibitem{Askariyan1} G.A. Askariyan, {\it Atomnaya Energiya} {\bf 3}, 152 (1957).  

\bibitem{Askariyan2} G.A. Askariyan et al., {\it Nucl. Inst. Meth.} {\bf 164}, 267 (1979).

\bibitem{Sulak} L. Sulak et al., {\it Nucl. Inst. Meth.} {\bf 161}, 203 (1979).

\bibitem{Hunter1} S.D. Hunter et al., {\it J. Acoust. Soc. Am.} {\bf 69}, 1557 (1981).

\bibitem{Hunter2} S.D. Hunter et al., {\it J. Acoust. Soc. Am.} {\bf 69}, 1563 (1981).
  
\bibitem{Albul} V.I. Albul et al., {\it Instr. Exp. Tech.} {\bf 44}, 327 (2001).

\bibitem{Karg} T. Karg et al., 1st International ARENA Workshop, Zeuthen (2005).

\end{thebibliography}
\end{document}